\documentclass[conference]{IEEEtran}

\usepackage{cite}

\usepackage{amsmath,amssymb,amsfonts}
\usepackage{graphicx}
\usepackage{array}
\usepackage{tabularx}
\newcolumntype{Y}{>{\raggedright\arraybackslash}X}
\usepackage{booktabs}
\usepackage{multirow}
\usepackage{makecell}

\usepackage{float}
\usepackage{placeins}

\usepackage{enumitem}

\usepackage{url}

\usepackage{algorithmicx}
\usepackage{algpseudocode}

\usepackage[ruled]{algorithm2e}

\newif\ifblind
\blindfalse   

\title{FIRCE: A Framework for Intrusion Response and Conformal Evaluation}

\ifblind
    \author{
    \IEEEauthorblockN{Anonymous Submission}
    \IEEEauthorblockA{Anonymous Institution}
    }
\else
    \author{
    \IEEEauthorblockN{Seth Barrett, Lin Li, Gokila Dorai}
    \IEEEauthorblockA{
    Augusta University, USA \\
    \{sebarrett, lli1, gdorai\}@augusta.edu
    }
    \and
    \IEEEauthorblockN{Swarnamugi Rajaganapathy}
    \IEEEauthorblockA{
    DFAIR Lab \\
    swarnamugi@dfairlab.com
    }
    }
\fi

\begin{document}

\maketitle
\begin{abstract}
Machine learning-based intrusion detection systems deployed in real-world environments frequently suffer from model degradation due to concept drift, where changes in traffic patterns invalidate training assumptions. To address this, we present FIRCE, a \textit{Framework for Intrusion Response and Conformal Evaluation} that augments supervised IDS classifiers with conformal evaluation-based uncertainty quantification and drift detection. FIRCE supports four conformal evaluation strategies: Inductive, Cross, Approximate Transductive, and our proposed Approximate Cross-Conformal Evaluator, which achieves robust performance with minimal calibration overhead. FIRCE also introduces an adaptive chunking mechanism that dynamically adjusts evaluation granularity in response to stream volatility, improving drift responsiveness while preserving computational efficiency. Using a custom IoT testbed of 10 commercial devices and time-series network captures under simulated attack and drift conditions, we demonstrate FIRCE’s ability to detect distributional shifts and trigger model retraining. We additionally benchmark FIRCE on the CICIDS2018 and UNSW-NB15 datasets to validate its generalizability. Experimental results show that conformal evaluation-based drift detection, combined with adaptive chunking, enables an efficient and robust response to evolving threats. 
\end{abstract}


\section{Introduction}
As Internet of Things (IoT) devices become increasingly integrated into critical infrastructures and everyday environments, their exposure to cyber threats grows proportionally. Intrusion Detection Systems (IDS) play a vital role in identifying unauthorized or anomalous behavior within these networks. However, most IDS models are trained in static environments and subsequently deployed in highly dynamic contexts. As traffic patterns evolve, due to firmware updates, new user behavior, or emerging attacks, models experience concept drift, where the underlying data distribution changes in ways that invalidate earlier assumptions~\cite{gama2014survey}. This leads to performance degradation and increased false negatives in identifying malicious activity.

To address this challenge, recent research has focused on uncertainty-aware and adaptive learning approaches that trigger retraining or rejection of unreliable predictions. Among these, conformal evaluation (CE) techniques stand out for their ability to quantify model confidence through calibrated p-values without requiring labeled data at test time. CE methods offer statistically grounded mechanisms for rejecting low-confidence predictions, making them ideal for drift detection in streaming, real-time environments such as IoT security monitoring~\cite{vovk2005algorithmic,shafer2008tutorial,angelopoulos2023conformal,jordaney2017transcend,barbero2022transcending}. Complementary lines of work explore concept-drift detection and response via explanation-aware filtering and pseudo-labeling~\cite{cade,adapt}, lifelong/unsupervised IDS pipelines under streaming shift~\cite{metanoia2025lifelong,le2021anomaly}, active/adaptive retraining under imbalance~\cite{gupta2025generative}, and adaptive window-sizing strategies for changing traffic regimes~\cite{BALDINI2022108923,baena2006early}. However, in practice these approaches still face calibration fragility under shift, costly or ill-timed retraining, error-amplifying pseudo-labels, and window-size sensitivity.

In this paper, we present FIRCE, a \textit{Framework for Intrusion Response and Conformal Evaluation}, designed to extend supervised IDS models with real-time drift detection and adaptive retraining. FIRCE incorporates a modular suite of conformal evaluators, including Inductive (ICE), Cross (CCE), Approximate Transductive (Approx-TCE), and our proposed Approximate Cross-Conformal Evaluator (Approx-CCE), which combines statistical rigor with low calibration overhead. In addition, FIRCE introduces an Adaptive Chunking Controller to dynamically adjust the number of flows processed in each simulation window, improving response to both stable and volatile traffic conditions.
\ifblind
\else
  The complete code-base and datasets are available for reproducibility~\cite{xseciot2025}.
\fi

\noindent Our contributions are as follows:
\begin{itemize}
    \item We develop ACC to adjust processing granularity based on recent drift activity, enabling more responsive and efficient real-time evaluation.
    \item We design and implement Approx-CCE, a novel conformal evaluation strategy that preserves the statistical strengths of CCE while requiring only a single model training pass.
    \item We adapt Transcendent’s~\cite{barbero2022transcending} conformal evaluators to a multilayer perceptron backbone, preserving calibration validity while improving throughput and enabling batched, GPU-friendly inference.
    \item We build and evaluate FIRCE on an IoT testbed containing ten consumer-grade devices under simulated attack and drift scenarios. Using time-series traffic captures and automated attack scheduling, we demonstrate that FIRCE can detect drift and trigger retraining with significantly improved strength compared to static classifiers.
\end{itemize}

\begin{figure*}[!t]
    \centering
    \includegraphics[width=\textwidth]{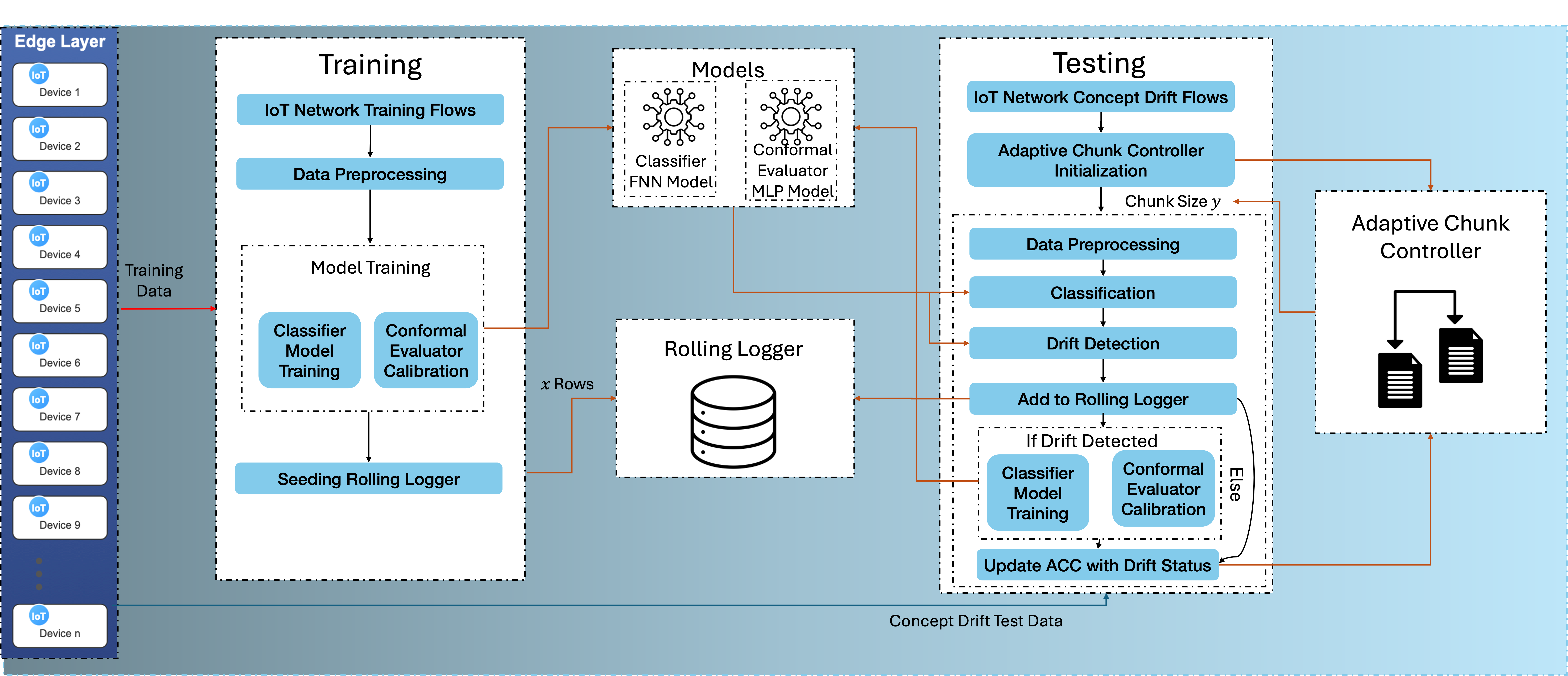}
    \caption{System-level overview of the FIRCE simulation pipeline. Flow chunks are streamed incrementally, evaluated using CE, and logged. Upon drift detection, retraining and recalibration are triggered from a rolling history.}
    \label{fig:ce-sim-framework}
\end{figure*}
The remainder of this paper is organized as follows: Section~\ref{sec:related-work} reviews related work on conformal evaluation, adaptive retraining and adaptive window sizes. Section~\ref{sec:methodology} details our methodology: data collection and preprocessing, baseline classifier configuration, framework design (CE integration, streaming simulation, and adaptive chunking), and design justifications. Section~\ref{sec:exp-protocol} discusses our experimental protocol and data pipeline. In Section~\ref{sec:results}, we discuss our experimental results and system performance. Section~\ref{sec:limitations} outlines key limitations and directions for future work. We conclude in Section~\ref{sec:conclusion}.

\section{Related Work}
\label{sec:related-work}

Adapting intrusion detection systems to non-stationary environments has been widely studied through drift detection, retraining strategies, and adaptive processing mechanisms. We summarize the most relevant directions below.

\subsection{Drift Detection and Adaptation in IDS}

CE provides a principled framework for uncertainty-aware drift detection by quantifying prediction reliability through nonconformity scores and p-values~\cite{vovk2005algorithmic, shafer2008tutorial}. Unlike conformal predictors, CE operates at the level of individual predictions, enabling low-confidence outputs to be rejected as potential indicators of distribution shift. Frameworks such as Transcend and Transcending Transcend demonstrate that conformal p-values can effectively identify violations of i.i.d. assumptions in security settings, providing statistically grounded drift signals~\cite{jordaney2017transcend,barbero2022transcending}. Compared to unsupervised anomaly detection methods~\cite{le2021anomaly}, CE separates prediction from uncertainty estimation, allowing more precise and interpretable drift detection with formal guarantees.

Drift adaptation in IDS broadly follows three directions: (i) score-based approaches that detect drift using conformity or p-value signals and trigger recalibration or retraining~\cite{jordaney2017transcend,barbero2022transcending}; (ii) confidence-driven pipelines that leverage pseudo-labeling, active learning, or augmentation to reduce labeling cost during updates~\cite{adapt,gupta2025generative}; and (iii) representation-centric methods that learn drift-resilient feature embeddings to enable lightweight adaptation~\cite{ReCDA,metanoia2025lifelong,soltani2023multi}. FIRCE primarily follows the first approach, using conformal evaluation for drift detection, while remaining compatible with confidence-based and representation-based updates.

\newcommand{\cmark}{\checkmark}
\newcommand{\xmark}{\(\times\)}

\begin{table*}[!t]
\caption{ML Models Used to Compute NCMs}
\label{tab:ncm-models}
\centering
\scriptsize
\setlength{\tabcolsep}{3pt}
\renewcommand{\arraystretch}{1.15}

\begin{tabularx}{\linewidth}{@{}p{0.16\linewidth}YYYYYY@{}}
\toprule
\textbf{Model} &
\makecell[l]{\textbf{I1}\\Prob.-based\\NCM $(1-p_{\hat y})$} &
\makecell[l]{\textbf{I2}\\Post-hoc\\calib.} &
\makecell[l]{\textbf{I3}\\Fast\\(re)train} &
\makecell[l]{\textbf{I4}\\Stable NCM\\(after calib.)} &
\makecell[l]{\textbf{I5}\\Good for\\tabular flows} &
\makecell[l]{\textbf{I6}\\Temp. scaling\\suffices} \\
\midrule
Linear / Kernel SVM
  & \cmark\footnotesize~(via Platt/iso)
  & \cmark
  & \makecell[l]{\small linear: \cmark\\\small kernel: \xmark}
  & \cmark\footnotesize~(margins)
  & \cmark
  & \xmark\footnotesize~(needs $A,B$ or iso) \\
Logistic Regression
  & \cmark & \cmark & \cmark & \cmark & \cmark & \xmark \\
Random Forest
  & \cmark\footnotesize~(vote probs)
  & \cmark & \cmark & \cmark & \cmark & \xmark \\
Gradient Boosting / XGBoost
  & \cmark\footnotesize~(\texttt{softprob})
  & \cmark & \cmark & \cmark & \cmark & \xmark \\
k-NN
  & \cmark\footnotesize~(freq.)
  & \small mixed
  & \xmark\footnotesize~(scale)
  & \small data-dependent
  & \small depends
  & \xmark \\
Naive Bayes
  & \cmark
  & \small mixed
  & \cmark
  & \small data-dependent
  & \cmark
  & \xmark \\
MLP
  & \textbf{\cmark}
  & \textbf{\cmark}\footnotesize~(temp.\ scaling)
  & \textbf{\cmark}\footnotesize~(compact)
  & \textbf{\cmark}\footnotesize~(with reg.)
  & \textbf{\cmark}
  & \cmark \\
\bottomrule
\end{tabularx}

\vspace{3pt}
\noindent\footnotesize
A \cmark\ indicates the model typically supports the criterion out-of-the-box or via a standard recipe; \xmark\ indicates it is uncommon or burdensome in practice.
\end{table*}

\subsection{Adaptability in Streaming IDS}

Adaptive processing mechanisms are commonly used to balance responsiveness and efficiency under concept drift. Methods such as ADWIN dynamically adjust window sizes based on detected changes in data distribution~\cite{adwin}, while related work employs EWMA-style controllers and early drift detection rules to improve sensitivity to evolving patterns~\cite{spinosa2009novelty,baena2006early}. 

In IoT intrusion detection, adaptive strategies have been explored through lightweight aggregation techniques~\cite{flare} and fog-layer chunked processing pipelines~\cite{fire}, which demonstrate the importance of dynamically controlling processing granularity for real-time performance. Motivated by these approaches, FIRCE introduces an Adaptive Chunking Controller that adjusts evaluation granularity based on observed drift behavior, improving both detection responsiveness and computational efficiency.

\subsection{Models for Nonconformity Measures}
\label{sec:ncm-models}
Nonconformity measures (NCMs) quantify how atypical a sample is relative to calibration data and are typically derived from either class probabilities or decision margins. The choice of model impacts calibration efficiency, stability under drift, and retraining cost. Table~\ref{tab:ncm-models} summarizes common model families used for NCM computation and their practical trade-offs in streaming settings.

In this work, we adopt compact multilayer perceptrons (MLPs) due to their ability to produce probability-based NCMs directly, support efficient retraining, and maintain stable calibration under sliding-window updates. These properties make MLPs well-suited for real-time conformal evaluation on tabular network flow data.

\begin{algorithm}[ht]
\caption{Approx-CCE: Calibration}
\label{alg:approxcce}
\footnotesize
\KwIn{Classifier $f$ with \texttt{fit}, \texttt{predict}, \texttt{predict\_proba}; data $(X,Y)$; folds $k$}
$\hat{f} \gets \texttt{clone}(f)$\;
$\hat{f}.\texttt{fit}(X,Y)$\;
Build stratified, shuffled $k$-folds $\{\mathcal{I}^{(j)}_{\mathrm{cal}}\}_{j=1}^k$\;
Init per-class buffers $S_c \gets [\,]$ for all $c$\;
\ForEach{$j=1..k$ \upshape{in parallel}}{
    $X^{(j)} \gets X[\mathcal{I}^{(j)}_{\mathrm{cal}}]$, $Y^{(j)} \gets Y[\mathcal{I}^{(j)}_{\mathrm{cal}}]$\;
    $P^{(j)} \gets \hat{f}.\texttt{predict\_proba}(X^{(j)})$\;
    \For{$i = 1..|Y^{(j)}|$}{
        $s_i \gets 1 - P^{(j)}_{i,\,Y^{(j)}_i}$\;
        Append $s_i$ to $S_{Y^{(j)}_i}$\;
    }
}
$\{\tau_c\} \gets \mathrm{Quantile}_{1-\alpha}(S_c)$ for all $c$\;
\Return{$\{S_c\}$, $\{\tau_c\}$}
\end{algorithm}


\begin{algorithm}[ht]
\caption{Approx-CCE Test-Time Prediction}
\label{alg:approxcce-test}
\KwIn{thresholds $\{\tau_c\}$, buffers $\{S_c\}$, model $\hat{f}$, sample $x$}
$P \gets \hat{f}.\texttt{predict\_proba}(x)$\;
$\hat{y} \gets \arg\max_c P[c]$\;
$s \gets 1 - P[\hat{y}]$\;
$p \gets \dfrac{1 + \sum_{s' \in S_{\hat{y}}} \mathbb{I}[s' \ge s]}{|S_{\hat{y}}| + 1}$ \tcp*{smoothed ECDF}
\eIf{$s > \tau_{\hat{y}}$}{
    \Return \texttt{Reject}\;
}{
    \Return \texttt{Accept} $(\hat{y},\, p)$\;
}
\end{algorithm}

\section{Methodology}
\label{sec:methodology}
This section outlines the data collection process, model configuration, CE framework, and simulation pipeline used to evaluate FIRCE (Figure~\ref{fig:ce-sim-framework}). We begin by describing the baseline classifiers. Next, we present the CE framework, including our Approx-CCE, and introduce the streaming simulation pipeline that mimics real-time deployment. We then justify key design choices and detail the ACC, followed by implementation notes on the MLP used within the CE components.

\subsection{Baseline Model}
\label{sec:Baseline}
We use a feedforward neural network (FNN) for network flow classification due to its strong performance and efficiency on tabular data. The model consists of two hidden layers (64, 32) with ReLU activations and dropout (0.3), trained using BCEWithLogitsLoss and Adam. Other models (DT, RF, SVM, XGBoost) are supported in our codebase but omitted here for brevity.

\textbf{Terminology and relationship to the CE model:}
We use FNN to refer to the classifier model reported in our results, and MLP to refer to the CE model described in Section~\ref{sec:ncm-models}. An MLP is a specific type of feedforward network; we adopt distinct names solely to avoid confusion between the classifier and the CE components.

\textbf{FNN classifier Parameters:}
The classifier is implemented with the following configuration, mirroring our released code:
(1) two hidden layers with widths $(64, 32)$;
(2) ReLU activations after each hidden layer;
(3) Dropout with probability $p_{\text{drop}}=0.3$ after each hidden layer;
(4) a single-logit output trained with BCEWithLogitsLoss and optimized with Adam (default learning rate $10^{-3}$);
(5) sigmoid applied to logits at inference to obtain $P(\text{attack})$;
(6) training and inference on CPU or GPU depending on availability.
This compact architecture was chosen to balance accuracy and runtime on structured, engineered flow features, consistent with recent findings on efficient tabular FNNs \cite{gorishniy2021revisiting, shwartz2022tabular}.

\subsection{Framework Design}
To reliably and efficiently detect concept drift and quantify prediction uncertainty in real time, FIRCE integrates a suite of CE techniques tailored for high-throughput network traffic scenarios.

\subsubsection{Why Conformal Evaluation for Drift Detection}
Conformal evaluation is uniquely suited to drift detection in network security contexts due to its statistical rigor, model-agnostic design, and real-time applicability. Key advantages include:

\begin{itemize}
    \item \textbf{No ground truth needed at test time:} CEs operate solely on model confidence, allowing drift detection before labels are available.
    \item \textbf{Per-class calibration:} Class-conditional thresholds improve sensitivity to nuanced distributional changes in specific traffic types.
    \item \textbf{Classifier independence:} Since CEs rely on prediction probabilities rather than internal model structure, they can be integrated with any probabilistic classifier.
\end{itemize}

These properties make CEs ideal for detecting subtle behavioral shifts in IoT network flows, where rapid reaction to threats is critical.

\subsubsection{Baseline Conformal Evaluators in FIRCE}
\label{sec:firce-ce-baselines}
FIRCE incorporates standard conformal evaluation methods, including ICE, CCE, and Approx-TCE~\cite{barbero2022transcending}. These methods differ primarily in calibration strategy and computational cost: ICE uses a single calibration split, CCE aggregates across multiple folds at higher cost, and Approx-TCE reduces transductive complexity through cached predictions. These baselines provide reference points for evaluating efficiency and performance trade-offs in streaming settings.


\subsubsection{Our Proposed Evaluator: Approx-CCE}
We propose Approx-CCE, a lightweight approximation of CCE designed for streaming environments. Instead of training $k$ models, Approx-CCE trains a single shared model and performs fold-based calibration across disjoint partitions. This preserves the statistical structure of cross-conformal evaluation while eliminating repeated training.

Calibration is performed in a single pass over the data to compute nonconformity scores and class-conditional thresholds, yielding $\mathcal{O}(n)$ complexity. At test time, p-values are computed via empirical distributions with constant-time lookup. This design enables efficient recalibration under frequent drift while maintaining compatibility with standard CE formulations. Algorithms for Approx-CCE calibration and test-time prediction can be found in algorithms ~\ref{alg:approxcce} and ~\ref{alg:approxcce-test}.

\subsubsection{Streaming Simulation}
FIRCE operates in a streaming setting using a fixed-size rolling calibration buffer. The buffer is initialized from training data and continuously updated with recent samples, ensuring that retraining and recalibration reflect current traffic behavior.

Incoming flows are processed in chunks. For each chunk, predictions and conformal p-values are computed, and drift is detected based on low-confidence outputs. When drift is detected, the model is retrained and the CE recalibrated using the most recent buffer contents. This design enables continuous adaptation under evolving traffic conditions.

\subsection{Framework Justification}
FIRCE is built for real-time operation under evolving traffic. First, it processes flows in controllable windows so calibration has sufficient statistical signal. Second, it separates prediction from uncertainty via a CE backbone, enabling drift detection before labels arrive. Third, it keeps update costs low with a rolling buffer and lightweight post-hoc calibration. Concretely, all CEs use a compact MLP as the internal model to produce probability-native nonconformity scores ($1-p_{\hat y}$) with simple temperature scaling for stable $p$-values across updates. To avoid the fixed-window trade-off between overreaction and sluggishness, we add an  ACC that adjusts evaluation granularity based on recent drift activity. The next subsection details ACC’s design and impact on responsiveness and runtime; we then describe the MLP configuration and its role in streaming CE.

\begin{algorithm}[ht]
\caption{Adaptive Chunk Size Controller}
\label{alg:adaptivechunking}
\KwIn{Initial size $\chi_0$, min size $\chi_{\text{min}}$, max size $\chi_{\text{max}}$, decay $\lambda$, step $\Delta$}
Set $\chi \gets \chi_0$, $\text{EMA} \gets 0.0$, $t_{\text{drift}} \gets 0$, $t_{\text{chunks}} \gets 0$\;
\While{data stream is not exhausted}{
    Extract chunk of size $\chi$ from input\;
    Run CE drift detection on chunk\;
    \If{drift detected}{
        $t_{\text{drift}} \gets t_{\text{drift}} + 1$\;
    }
    $t_{\text{chunks}} \gets t_{\text{chunks}} + 1$\;
    Compute raw drift rate: $r \gets \frac{t_{\text{drift}}}{t_{\text{chunks}}}$\;
    Update EMA: $\text{EMA} \gets \lambda \cdot \text{EMA} + (1 - \lambda) \cdot r$\;
    \eIf{$\text{EMA} > 0.2$}{
        $\chi \gets \max(\chi_{\text{min}}, \lfloor \chi / 2 \rfloor)$\;
    }{
        \If{$\text{EMA} < 0.05$}{
            $\chi \gets \min(\chi_{\text{max}}, \chi + \Delta)$\;
        }
    }
    Append $\chi$ to chunk size trace\;
}
\end{algorithm}


\subsubsection{Adaptive Chunking}
To balance detection sensitivity and computational efficiency, FIRCE introduces an Adaptive Chunking Controller (ACC) that dynamically adjusts the number of flows processed per evaluation window. 

Small chunks improve responsiveness but increase retraining frequency, while large chunks reduce overhead but delay drift detection. ACC addresses this trade-off by adjusting chunk size based on recent drift activity using an exponential moving average (EMA) of drift rates. When drift frequency increases, chunk sizes decrease to improve responsiveness; during stable periods, chunk sizes increase to reduce computation.

This adaptive strategy enables FIRCE to maintain high detection performance while avoiding both excessive retraining and delayed drift response. The algorithm covering ACC can be found in algorithm ~\ref{alg:adaptivechunking}.

\begin{algorithm}[ht]
\caption{MLP \texttt{predict\_proba}}
\label{alg:mlp-predict-proba}
\footnotesize
\KwIn{$X \in \mathbb{R}^{N\times D}$ or $x\in\mathbb{R}^D$; batch size $B$}
If $X$ is 1D, reshape to $(1,D)$; ensure \texttt{float32}, C-contiguous\;
Split $X$ into mini-batches $\{X_b\}$ of size $\le B$\;
\For{$X_b$}{
    $z_b \gets \texttt{forward}(X_b)$ \tcp*{single logit per row}
    $p_b \gets \sigma(z_b)$ \tcp*{$\sigma(t)=1/(1+e^{-t})$}
    Append $\big[1-p_b,\; p_b\big]$ to list\;
}
\Return{$\texttt{stack}(\text{list}) \in \mathbb{R}^{N\times 2}$}
\end{algorithm}


\subsubsection{MLP for Conformal Evaluation}
\label{sec:mlps}
FIRCE uses a compact multilayer perceptron (MLP) to produce probability-based nonconformity scores ($1 - p_{\hat y}$), as seen in ~\ref{alg:mlp-predict-proba}. The model consists of three hidden layers (256, 128, 64) with GELU activations, LayerNorm, and dropout (0.2), trained using BCEWithLogitsLoss and Adam.

MLPs are selected due to their ability to produce calibrated probabilities efficiently, enabling lightweight temperature scaling and fast retraining under streaming conditions. Compared to SVM-based approaches, this avoids costly per-update calibration procedures while maintaining competitive performance on tabular network flow data.

\section{Experimental Protocol and Data Pipeline}
\label{sec:exp-protocol}
This section details how we (i) collected and labeled IoT traffic in a controlled environment, (ii) transformed packet captures into standardized flow features for model training, and (iii) configured streaming evaluation runs across in-lab and benchmark datasets. Together, these steps define a reproducible protocol for assessing drift detection, retraining, and recalibration behavior under realistic IoT traffic conditions.

\begin{table*}[!t]
\centering
\ifblind
\else
  \caption{Approx\mbox{-}CCE vs.\ CE variants across DFAIR and cross\mbox{-}dataset benchmarks}
\fi
\label{tab:headline-results}

\scriptsize
\setlength{\tabcolsep}{3pt}
\renewcommand{\arraystretch}{1.12}

\begin{tabularx}{\linewidth}{@{}p{0.16\linewidth}p{0.16\linewidth}
*{4}{>{\centering\arraybackslash}p{0.085\linewidth}}
>{\centering\arraybackslash}p{0.14\linewidth}
>{\centering\arraybackslash}p{0.10\linewidth}@{}}
\toprule
\textbf{Dataset} & \textbf{CE Type} & \textbf{CE Acc.} & \textbf{Prec.} & \textbf{Rec.} & \textbf{F1} & \textbf{Runtime (s)} & \textbf{\# Calibs} \\
\midrule

\ifblind
\else
  \multirow{4}{*}{\parbox{0.16\linewidth}{\raggedright DFAIR}}
\fi
  & ICE                 & 1.0000 & 0.9999 & 0.9999 & 0.9999  & 626.6639   & No Retrain \\
  & Approx-TCE          & 1.0000 & 1.0000 & 1.0000 & 1.0000  & 768.7549   & 4 \\
  & CCE                 & 1.0000 & 0.9998 & 0.9999 & 0.9998  & 785.3604   & 2 \\
  & \textbf{Approx-CCE} & \textbf{1.0000} & \textbf{0.9998} & \textbf{0.9999} & \textbf{0.99987} & \textbf{676.2752} & \textbf{2} \\
\midrule

\multirow{4}{*}{\parbox{0.16\linewidth}{\raggedright CICIDS2018}}
  & ICE                 & 0.9953 & 0.8990 & 0.9701 & 0.9288 & 5628.1554  & 2 \\
  & Approx-TCE          & 0.9976 & 0.9111 & 0.9852 & 0.9439 & 6637.4307  & 4 \\
  & CCE                 & 0.9950 & 0.9962 & 0.9667 & 0.9811 & 16584.1373 & 2 \\
  & \textbf{Approx-CCE} & \textbf{0.9952} & \textbf{0.9973} & \textbf{0.9667} & \textbf{0.9816} & \textbf{6020.9825} & \textbf{2} \\
\midrule

\multirow{4}{*}{\parbox{0.16\linewidth}{\raggedright NB15}}
  & ICE                 & 0.9988 & 0.9897 & 0.9962 & 0.9929 & 600.9195   & No Retrain \\
  & Approx-TCE          & 0.9991 & 0.9939 & 0.9961 & 0.9950 & 720.1210   & 4 \\
  & CCE                 & 0.9975 & 0.9881 & 0.9842 & 0.9861 & 1792.7051  & 7 \\
  & \textbf{Approx-CCE} & \textbf{0.9980} & \textbf{0.9892} & \textbf{0.9885} & \textbf{0.9889} & \textbf{717.1803} & \textbf{4} \\
\bottomrule
\end{tabularx}
\end{table*}

\subsection{IoT Traffic Capture, Labeling, and Flow Preprocessing}
\label{sec:DataCollection}
To develop and evaluate our intrusion detection framework, we collected labeled IoT network traffic in a controlled laboratory environment. Our experimental network included a TP-Link TL-WR541N router running OpenWrt firmware and ten commercial off-the-shelf IoT devices spanning smart home, security, and entertainment domains. Each device was assigned a static IP address and interacted with via its corresponding mobile application or voice assistant, enabling consistent behavioral profiling. 


\ifblind
\else
    We conducted two sets of data captures: (1) DFAIR-Train, containing benign traffic and attack simulations for model training, and (2) DFAIR-Drift, containing additional captures under concept drift conditions to evaluate model adaptation.
\fi
All traffic captures were conducted using \texttt{tcpdump} on a Debian Linux laptop directly connected to the IoT subnet. For each device, an 8-hour traffic capture was taken. During this time, a series of network-layer attacks were periodically launched from the same Debian laptop, including TCP SYN flood, XMAS tree flood, UDP flood, HTTP flood, and, in the case of drift captures, a HULK HTTP flood.

To ensure consistency across devices and maintain balanced exposure to each attack type, attacks were executed using predefined intervals. For every device, each attack type was launched three times across the 8-hour window. A safe cooldown period was preserved in the final 15 minutes to avoid trailing attack packets affecting benign classification.
\ifblind
\else
    All attack timestamps were logged for ground-truth labeling, and the capture and attack automation scripts are publicly available in the CAPEX Capture repository~\cite{capex2024capture}.
\fi

Following raw PCAP collection, flows were extracted using CICFlowMeter, and were labeled based on attack logs. Before training, all flow features are standardized using a StandardScaler, and only numeric columns are retained.
For binary classification, the label is derived from the \texttt{Label} or \texttt{BinLabel} column, mapping ``Benign'' to 0 and all others to 1.

The training routines are implemented in the \texttt{ce\_model\_training.py} module, which handles preprocessing, model instantiation, training, and artifact persistence (i.e., \texttt{.pkl} or \texttt{.pt} files). After training, each model is evaluated on the full training dataset. Metrics including accuracy, precision, recall, and F1 score are computed and logged to assess model fit prior to simulation. These metrics are retained for both Feedforward Neural Network (FNN) classifier (discussed in \ref{sec:Baseline}) and the internal MLP model (discussed in \ref{sec:mlps}) used by CEs to ensure consistency and traceability across the detection pipeline. This training and evaluation process is automatically triggered during simulation initialization if no saved model artifacts are found.

\subsection{Streaming Evaluation Configuration and Benchmark Transfer}
\label{sec:stream-config}
We next describe how the trained classifier and conformal evaluators are exercised in our streaming simulation, and how we define training, calibration, and evaluation splits across in-lab and benchmark datasets. Unless noted, classifier metrics are computed after each retrain using the latest labeled data available in the rolling window; CE metrics are computed on the same stream, with p-values and thresholds recalibrated on the sliding calibration buffer.

\ifblind
\else
    For DFAIR (our main run), both the classifier and the CE are trained/calibrated on our in-lab dataset, DFAIR-Train (Sec.~\ref{sec:DataCollection}) and evaluated on its concept-drift variant, DFAIR-Drift, collected on the same network and devices but with new attacks (also Sec.~\ref{sec:DataCollection}). For the UNSW benchmark run, the classifier/CE are trained/calibrated on UNSW-NB15~\cite{unsw-nb15} and evaluated on the DFAIR-Drift dataset. For the CICIDS2018 benchmark run, the classifier/CE are trained/calibrated on CICIDS2018~\cite{cicids2018} and evaluated on the DFAIR-Drift.
\fi

Differences in retraining and recalibration counts primarily reflect dataset characteristics rather than our feature subset. First, CICIDS2018 is organized into scenario-batched days (e.g., DoS/DDoS day, Web-attacks day), yielding long semi-stationary segments that ACC can span with larger chunks and fewer updates~\cite{unb2018Datasets,opendataRealisticCyber}. Second, UNSW-NB15 blends live traffic with IXIA PerfectStorm--synthesized streams across nine attack families; such mixes change rates and patterns more abruptly than enterprise-style traces and typically induce more recalibration when transferred to
\ifblind
\else
    DFAIR
\fi
~\cite{unswUNSWNB15Dataset,unsw-nb15}. Third, even with the same field names, CICIDS2018 flows are produced by CICFlowMeter-V3 whereas UNSW-NB15 features come from Argus/Bro; differences in biflow definitions and timers shift the distribution of identically named statistics (e.g., \texttt{flow\_duration}, \texttt{fwd\_iat\_*}), affecting calibration stability~\cite{unbApplicationsResearch,unswUNSWNB15Dataset}. Finally, multiple studies report higher separability on CICIDS201x (often $>95$--$99\%$ accuracy with standard models), which corresponds to smoother posteriors and fewer threshold adjustments under ACC~\cite{leevy2020survey,songma2023optimizing}.

\section{Results}
\label{sec:results}
\ifblind
\else
    We evaluate FIRCE on an in-lab IoT stream (DFAIR-Train + DFAIR-Drift) and on transfer settings where models are trained/calibrated on UNSW-NB15 or CICIDS2018 and evaluated on DFAIR-Drift, tracking Accuracy/Precision/Recall/F1, calibration behavior, and stability under drift under streaming updates. 
\fi
We evaluate whether Approx-CCE maintains CCE-level performance with lower cost, and whether ACC reduces calibration overhead without degrading detection quality.

\subsection{Approx-CCE Findings}
Table~\ref{tab:headline-results} shows that Approx-CCE matches CCE in accuracy and F1 across all settings while consistently reducing runtime. ICE occasionally fails to trigger retraining due to conservative calibration, while Approx-TCE remains slower due to its multi-fold structure. These results highlight Approx-CCE as the most practical CE variant for streaming IDS.

\ifblind
\else
    \subsubsection{Main Finding on DFAIR-Train + DFAIR-Drift}
    On DFAIR, Approx-CCE matches CCE in accuracy and F1 while reducing runtime by 13.9\% (676s vs.\ 785s) with identical calibration counts.
\fi

\begin{table*}[!t]
\caption{Adaptive Chunking Controller with Approx-CCE Results}
\label{tab:chunking-results}
\centering
\scriptsize
\setlength{\tabcolsep}{3pt}
\renewcommand{\arraystretch}{1.08}

\begin{tabularx}{\linewidth}{@{}
p{0.16\linewidth}
p{0.12\linewidth}
p{0.13\linewidth}
p{0.11\linewidth}
p{0.11\linewidth}
p{0.11\linewidth}
p{0.11\linewidth}
p{0.12\linewidth}
@{}}
\toprule
\textbf{Dataset} &
\textbf{Chunk Size} &
\textbf{Num.\ Calibrations} &
\textbf{CE Accuracy} &
\textbf{CE Precision} &
\textbf{CE Recall} &
\textbf{CE F1 Score} &
\textbf{Runtime (s)} \\
\midrule

\ifblind
\else
  \multirow{11}{*}{\parbox{0.16\linewidth}{\raggedright DFAIR}}
\fi
  & 100          & 8         & 1.0000 & 0.9997 & 1.0000 & 0.9998 &   863.6948 \\
  & 75           & 12        & 1.0000 & 0.9997 & 1.0000 & 0.9998 &  1115.1039 \\
  & 50           & 11        & 0.9993 & 0.9991 & 0.9957 & 0.9973 &  1043.4760 \\
  & 25           & 14        & 1.0000 & 0.9996 & 1.0000 & 0.9998 &  1277.0447 \\
  & 15           & 36        & 0.9995 & 0.9990 & 0.9970 & 0.9980 &  3073.3729 \\
  & 10           & 27        & 0.9997 & 0.9995 & 0.9983 & 0.9989 &  2864.6938 \\
  & 5            & 41        & 0.9995 & 0.9996 & 0.9964 & 0.9980 &  3426.8487 \\
  & 1            & 172       & 0.9996 & 0.9996 & 0.9967 & 0.9981 & 13788.0775 \\
  & \textbf{Adaptive} & \textbf{2} & \textbf{1.0000} & \textbf{0.9998} & \textbf{0.9999} & \textbf{0.9998} & \textbf{785.3604} \\
\midrule

\multirow{11}{*}{\parbox{0.16\linewidth}{\raggedright UNSW-NB15}}
  & 100          & 60        & 0.9966 & 0.9846 & 0.9776 & 0.9811 &   4923.7553 \\
  & 75           & 45        & 0.9967 & 0.9856 & 0.9780 & 0.9818 &   3864.0136 \\
  & 50           & 106       & 0.9966 & 0.9844 & 0.9772 & 0.9808 &   7759.0822 \\
  & 25           & 191       & 0.9965 & 0.9841 & 0.9771 & 0.9806 &  12957.9663 \\
  & 15           & 142       & 0.9969 & 0.9869 & 0.9783 & 0.9826 &   9963.9503 \\
  & 10           & 414       & 0.9966 & 0.9848 & 0.9773 & 0.9810 &  28020.7902 \\
  & 5            & 570       & 0.9968 & 0.9860 & 0.9782 & 0.9821 &  36997.5803 \\
  & 1            & 3133      & 0.9967 & 0.9854 & 0.9777 & 0.9815 & 204751.6373 \\
  & \textbf{Adaptive} & \textbf{7} & \textbf{0.9975} & \textbf{0.9881} & \textbf{0.9842} & \textbf{0.9861} & \textbf{1792.7051} \\
\midrule

\multirow{11}{*}{\parbox{0.16\linewidth}{\raggedright CICIDS2018}}
  & 100          & 2         & 0.9951 & 0.9968 & 0.9666 & 0.9813 &  16783.3195 \\
  & 75           & 2         & 0.9951 & 0.9968 & 0.9666 & 0.9813 &  17854.7042 \\
  & 50           & 2         & 0.9951 & 0.9967 & 0.9666 & 0.9813 &  15941.0160 \\
  & 25           & 2         & 0.9951 & 0.9951 & 0.9969 & 0.9814 &  16294.3048 \\
  & 15           & 3         & 0.9963 & 0.9977 & 0.9746 & 0.9859 &  16341.1437 \\
  & 10           & 2         & 0.9951 & 0.9969 & 0.9666 & 0.9814 &  15823.8786 \\
  & 5            & 2         & 0.9951 & 0.9968 & 0.9667 & 0.9814 &  16232.5535 \\
  & 1            & 3         & 0.9963 & 0.9978 & 0.9749 & 0.9861 &  17646.2401 \\
  & \textbf{Adaptive} & \textbf{2} & \textbf{0.9950} & \textbf{0.9962} & \textbf{0.9667} & \textbf{0.9811} & \textbf{16584.1373} \\
\bottomrule
\end{tabularx}
\end{table*}

\subsubsection{Cross-Dataset (Domain-Shift) Benchmarks}
Under domain shift, Approx-CCE continues to match CCE while reducing runtime by 60–64\% across CICIDS2018 and UNSW-NB15, with equal or fewer recalibrations.

\ifblind
\else
\subsubsection{On DFAIR’s Near-Perfect Scores}
    The near-ceiling performance on DFAIR reflects an in-domain drift setting where feature distributions remain aligned despite new attacks. In contrast, cross-dataset evaluations introduce covariate and prior shift, leading to slightly lower but still strong performance. In both cases, Approx-CCE maintains parity with CCE while enabling calibrated rejection under drift.
\fi

\subsubsection{Takeaway}
Across all datasets, Approx-CCE matches CCE within $3\times 10^{-4}$ on accuracy and within $10^{-3}$ on F1 while being computationally cheaper. Under domain shift (UNSW-NB15 and CICIDS2018 benchmarks), both CE variants yield higher F1 than classifier-only, indicating more reliable decisions in the presence of drift. On 
\ifblind
\else
    DFAIR’s
\fi
in-domain drift, CE maintains parity with the classifier while enabling principled rejection/alerting.

\subsection{Adaptive Chunking Controller Findings}
\label{sec:acc-results}
The ACC targets the core IDS requirement of timely model upkeep: it schedules calibrations on-the-fly to control time cost while preserving detection quality. In contrast to fixed chunking, where too-small chunks trigger calibration explosions and too-large chunks delay responsiveness, ACC expands chunks during stable periods and contracts them around drift, yielding fewer total calibrations and bounded runtime without sacrificing Accuracy/Precision/Recall/F1 (Table~\ref{tab:chunking-results}). This aligns with operational guidance emphasizing rapid detection/containment in incident response~\cite{nist80061r3,verizon2025dbir}.

\ifblind
\else
    \begin{itemize}
        \item On \textit{DFAIR-Train} + \textit{DFAIR-Drift,} ACC achieves 1.0000 / 0.9998 / 0.9999 / 0.9998 (Accuracy/Precision/Recall/F1) with only 2 calibrations in 785.36\,s. Fixed small chunks attain similar top-line metrics but at substantially higher calibration counts and runtime (e.g., chunk $1$: 172 calibrations, $13{,}788.08$\,s). Larger fixed chunks (e.g., $1000$, $500$) are fast (502.51–570.67\,s) but exhibit reduced recall/F1 (e.g., $0.9832/0.9913$, $0.9874/0.9934$), illustrating the trade-off that ACC mitigates without manual tuning.

        \item On \textit{UNSW-NB15} + \textit{DFAIR-Drift,} ACC maintains high utility with 0.9975 / 0.9881 / 0.9842 / 0.9861 in 1{,}792.71\,s using 7 calibrations. Relative to tiny fixed chunks, ACC avoids overhead (e.g., chunk $1$: 3{,}133 calibrations, $204{,}751.64$\,s) while matching or exceeding the mid-range fixed-chunk scores (e.g., chunk $100$: F1 $0.9811$ vs.\ ACC $0.9861$). Thus ACC preserves accuracy under domain shift while capping calibration frequency and elapsed time.

        \item On \textit{CICIDS2018} + \textit{DFAIR-Drift,} ACC achieves 0.9950 / 0.9962 / 0.9667 / 0.9811 with 2 calibrations in 16{,}584.14\,s, statistically on par with fixed-chunk results but without the need to pre-select a global chunk size. Here the benefit is primarily operational: ACC removes chunk-size guesswork and prevents calibration spikes if drift frequency increases.
    \end{itemize}
\fi

\textbf{Takeaway:}
ACC maintains detection quality while significantly reducing calibration overhead, enabling stable and efficient streaming IDS operation.

\section{Limitations and Future Work}
\label{sec:limitations}
\textbf{Model scope and labeling:}
This study focuses on binary intrusion detection (benign vs. malicious). Extending FIRCE to multi-class settings and to open-world detection of previously unseen attack families remains future work. A practical next step is to couple CE rejections with human-in-the-loop or LLM-assisted labeling workflows that propose candidate class names, rationales, and cluster assignments for novel traffic, reducing turnaround time for incorporating new classes.

\textbf{Classifier retraining policy:}
Our pipeline retrains the classifier on recently labeled data from the rolling buffer. While CE-based drift detection and p-value rejection operate independently of label quality, the retraining step can inherit biases from imperfect labels, especially during sustained drift. Future work will examine more conservative retraining triggers, uncertainty-aware pseudo-label filtering, and active sampling strategies that target ambiguous flows before they influence the next training cycle.

\textbf{Dataset coverage and rolling buffer size:}
Large portions of UNSW-NB15 and CICIDS2018 lack attack labels in certain slices. To ensure that retraining windows contain attack traffic, we configured a large rolling logger. While this stabilizes training and calibration, it increases memory and may delay the influence of the most recent samples. Future work will quantify this trade-off and explore adaptive windowing, class-balanced replay, and label-efficient sampling that preserves minority attacks without inflating the buffer.

\textbf{Attack diversity:}
The present evaluation emphasizes DoS-style attacks. Understanding how conformal evaluation behaves with other network attack categories (e.g., reconnaissance, data exfiltration, botnet C2, and lateral movement) is an important next step. We plan to expand the attack mix and to analyze per-class coverage, rejection behavior, and time-to-detection under mixed, multi-vector adversarial campaigns.

\textbf{Portability and determinism:}
During development we encountered crashes tied to device-side random permutations on Apple M1 hardware with our PyTorch stack. To ensure reproducibility across devices and enable threaded Approx-CCE calibration, we adopted deterministic, CPU-side epoch shuffling via a NumPy RNG and added thread-safe inference in the CE model. This design favors stability over maximum throughput; future work will revisit fully device-side data loaders once the underlying issues are resolved in the toolchain.

\textbf{Scalability and Distributed deployment:} FIRCE assumes a centralized model trained and retrained on a single node. In large-scale IoT environments with many heterogeneous devices, a federated or hierarchical deployment may be necessary. Extending FIRCE to distributed settings while preserving conformal guarantees is a promising direction for future work.



\section{Conclusion}
\label{sec:conclusion}
We presented FIRCE, a streaming intrusion detection framework built on conformal evaluation, validated on a custom IoT testbed of ten commercial devices as well as the CICIDS2018 and UNSW-NB15 benchmark datasets. Approx-CCE preserves the detection performance of CCE while significantly reducing calibration cost, achieving parity within $3\times10^{-4}$ on accuracy and $10^{-3}$ on F1 across all evaluated settings. ACC dynamically adjusts evaluation granularity to balance responsiveness and efficiency under drift, eliminating the need for manual chunk-size tuning while avoiding both calibration explosions and delayed drift response, as evidenced by the results in Table~\ref{tab:chunking-results}.

Across both in-domain and cross-dataset transfer settings, FIRCE maintains strong detection performance while reducing runtime and calibration overhead, demonstrating its practicality for real-time IDS deployment in dynamic IoT environments. Nonetheless, open challenges remain in threshold sensitivity, hard real-time feasibility, adversarial robustness, and scalability to distributed or federated deployments.

Future work will extend FIRCE to multi-class and open-world settings, incorporate LLM-assisted labeling workflows for novel attack families, and explore federated variants suitable for large-scale heterogeneous IoT infrastructures.

\ifblind
\else
  \section*{Acknowledgments}
  We would like to acknowledge AU TGS and GSGA for funding the publication of this work.
  Additionally, we acknowledge Bradley Boswell for his help and work on this project.
\fi

\bibliographystyle{splncs04}
\bibliography{main}

@inproceedings{barbero2022transcending,
	title        = {Transcending transcend: Revisiting malware classification in the presence of concept drift},
	author       = {Barbero, Federico and Pendlebury, Feargus and Pierazzi, Fabio and Cavallaro, Lorenzo},
	year         = 2022,
	booktitle    = {2022 IEEE Symposium on Security and Privacy (SP)},
	pages        = {805--823},
	organization = {IEEE}
}

@inproceedings{jordaney2017transcend,
	title        = {Transcend: Detecting concept drift in malware classification models},
	author       = {Jordaney, Roberto and Sharad, Kumar and Dash, Santanu K and Wang, Zhi and Papini, Davide and Nouretdinov, Ilia and Cavallaro, Lorenzo},
	year         = 2017,
	booktitle    = {26th USENIX security symposium (USENIX security 17)},
	pages        = {625--642}
}

@misc{xseciot2025,
	title        = {XSecIoT - FIRCE Backup Branch},
	author       = {Seth Barrett},
	year         = 2025,
	note         = {Accessed: 2025-06-30},
	howpublished = {\url{https://github.com/DFAIR-LAB-Augusta/XSecIoT/tree/FIRCE_bkp}}
}

@misc{capex2024capture,
	title        = {{CAPEX-Capture-for-Evaluation: IoT Attack and Baseline Data Capture Scripts}},
	author       = {Seth Barrett},
	year         = 2024,
	note         = {Accessed: 2025-06-30},
	howpublished = {\url{https://github.com/DFAIR-LAB-Augusta/CAPEX-Capture-for-Evaluation}}
}

@book{vovk2005algorithmic,
	title        = {Algorithmic learning in a random world},
	author       = {Vovk, Vladimir and Gammerman, Alexander and Shafer, Glenn},
	year         = 2005,
	publisher    = {Springer},
	volume       = 29
}

@article{shafer2008tutorial,
	title        = {A tutorial on conformal prediction.},
	author       = {Shafer, Glenn and Vovk, Vladimir},
	year         = 2008,
	journal      = {Journal of Machine Learning Research},
	volume       = 9,
	number       = 3
}

@article{le2021anomaly,
	title        = {Anomaly detection for insider threats using unsupervised ensembles},
	author       = {Le, Duc C and Zincir-Heywood, Nur},
	year         = 2021,
	journal      = {IEEE Transactions on Network and Service Management},
	publisher    = {IEEE},
	volume       = 18,
	number       = 2,
	pages        = {1152--1164}
}

@inproceedings{cade,
  title={$\{$CADE$\}$: Detecting and explaining concept drift samples for security applications},
  author={Yang, Limin and Guo, Wenbo and Hao, Qingying and Ciptadi, Arridhana and Ahmadzadeh, Ali and Xing, Xinyu and Wang, Gang},
  booktitle={30th USENIX Security Symposium (USENIX Security 21)},
  pages={2327--2344},
  year={2021}
}

@article{adapt,
  title={ADAPT: A Pseudo-labeling Approach to Combat Concept Drift in Malware Detection},
  author={Alam, Md Tanvirul and Piplai, Aritran and Rastogi, Nidhi},
  journal={arXiv preprint arXiv:2507.08597},
  year={2025}
}

@article{BALDINI2022108923,
title = {Online Distributed Denial of Service (DDoS) intrusion detection based on adaptive sliding window and morphological fractal dimension},
journal = {Computer Networks},
volume = {210},
pages = {108923},
year = {2022},
issn = {1389-1286},
doi = {https://doi.org/10.1016/j.comnet.2022.108923},
url = {https://www.sciencedirect.com/science/article/pii/S138912862200113X},
author = {Gianmarco Baldini and Irene Amerini}
}

@article{spinosa2009novelty,
  title={Novelty detection with application to data streams},
  author={Spinosa, Eduardo J and de Carvalho, Andr{\'e} Ponce de Leon F and Gama, Jo{\~a}o},
  journal={Intelligent Data Analysis},
  volume={13},
  number={3},
  pages={405--422},
  year={2009},
  publisher={SAGE Publications Sage UK: London, England}
}

@inproceedings{baena2006early,
  title={Early drift detection method},
  author={Baena-Garc{\i}a, Manuel and del Campo-{\'A}vila, Jos{\'e} and Fidalgo, Raul and Bifet, Albert and Gavalda, Ricard and Morales-Bueno, Rafael},
  booktitle={Fourth international workshop on knowledge discovery from data streams},
  volume={6},
  pages={77--86},
  year={2006}
}

@article{soltani2023multi,
  title={A multi-agent adaptive deep learning framework for online intrusion detection},
  author={Soltani, Mahdi and Khajavi, Khashayar and Jafari Siavoshani, Mahdi and Jahangir, Amir Hossein},
  journal={Cybersecurity},
  volume={7},
  number={1},
  pages={9},
  year={2024},
  publisher={Springer}
}

@article{gupta2025generative,
  title={Generative active adaptation for drifting and imbalanced network intrusion detection},
  author={Gupta, Ragini and Liu, Shinan and Zhang, Ruixiao and Hu, Xinyue and Kommaraju, Pranav and Wang, Xiaoyang and Benkraouda, Hadjer and Feamster, Nick and Nahrstedt, Klara},
  journal={arXiv preprint arXiv:2503.03022},
  year={2025}
}

@article{metanoia2025lifelong,
  title={METANOIA: A Lifelong Intrusion Detection and Investigation System for Mitigating Concept Drift},
  author={Ying, Jie and Zhu, Tiantian and Zheng, Aohan and Chen, Tieming and Lv, Mingqi and Chen, Yan},
  journal={arXiv preprint arXiv:2501.00438},
  year={2024}
}

@inproceedings{unsw-nb15,
	title        = {UNSW-NB15: a comprehensive data set for network intrusion detection systems (UNSW-NB15 network data set)},
	author       = {Moustafa, Nour and Slay, Jill},
	year         = 2015,
	booktitle    = {2015 Military Communications and Information Systems Conference (MilCIS)},
	volume       = {},
	number       = {},
	pages        = {1--6},
	doi          = {10.1109/MilCIS.2015.7348942}
}

@article{cicids2018,
  title={Toward generating a new intrusion detection dataset and intrusion traffic characterization.},
  author={Sharafaldin, Iman and Lashkari, Arash Habibi and Ghorbani, Ali A and others},
  journal={ICISSp},
  volume={1},
  number={2018},
  pages={108--116},
  year={2018}
}

@article{angelopoulos2023conformal,
  title={Conformal prediction: A gentle introduction},
  author={Angelopoulos, Anastasios N and Bates, Stephen and others},
  journal={Foundations and trends{\textregistered} in machine learning},
  volume={16},
  number={4},
  pages={494--591},
  year={2023},
  publisher={Now Publishers, Inc.}
}

@article{ReCDA,
  title={Self-Supervised Adaptation Method to Concept Drift for Network Intrusion Detection},
  author={Yang, Shuo and Zheng, Xinran and Li, Jinze and Xu, Jinfeng and Zhang, Xinchen and Ngai, Edith CH},
  journal={IEEE Transactions on Dependable and Secure Computing},
  year={2025},
  publisher={IEEE}
}

@article{gorishniy2021revisiting,
  title={Revisiting deep learning models for tabular data},
  author={Gorishniy, Yury and Rubachev, Ivan and Khrulkov, Valentin and Babenko, Artem},
  journal={Advances in neural information processing systems},
  volume={34},
  pages={18932--18943},
  year={2021}
}

@article{shwartz2022tabular,
  title={Tabular data: Deep learning is not all you need},
  author={Shwartz-Ziv, Ravid and Armon, Amitai},
  journal={Information Fusion},
  volume={81},
  pages={84--90},
  year={2022},
  publisher={Elsevier}
}

@article{flare,
  title={FLARE: Feature-based Lightweight Aggregation for Robust Evaluation of IoT Intrusion Detection},
  author={Boswell, Bradley and Barrett, Seth and Rajaganapathy, Swarnamugi and Dorai, Gokila},
  journal={arXiv preprint arXiv:2504.15375},
  year={2025}
}

@inproceedings{fire,
  title={FIRE: Fog-Based Intrusion Detection Framework for Real-Time Security in IoT Environments},
  author={Boswell, Bradley and Dorai, Gokila and Barrett, Seth and Rajaganapathy, Swarnamugi and Li, Lin},
  booktitle={Proceedings of the Future Technologies Conference},
  pages={209--226},
  year={2025},
  organization={Springer}
}

@inproceedings{adwin,
  title={Learning from time-changing data with adaptive windowing},
  author={Bifet, Albert and Gavalda, Ricard},
  booktitle={Proceedings of the 2007 SIAM international conference on data mining},
  pages={443--448},
  year={2007},
  organization={SIAM}
}

@techreport{nist80061r3,
  author       = {Alex Nelson and Sanjay Rekhi and Murugiah Souppaya and Karen Scarfone},
  title        = {Incident Response Recommendations and Considerations for Cybersecurity Risk Management: A CSF 2.0 Community Profile},
  institution  = {National Institute of Standards and Technology},
  type         = {NIST Special Publication},
  number       = {NIST SP 800-61r3},
  address      = {Gaithersburg, MD},
  year         = {2025},
  month        = apr,
  doi          = {10.6028/NIST.SP.800-61r3},
  url          = {https://doi.org/10.6028/NIST.SP.800-61r3}
}

@techreport{verizon2025dbir,
  author       = {{Verizon}},
  title        = {2025 Data Breach Investigations Report},
  institution  = {Verizon},
  address      = {Basking Ridge, NJ, USA},
  type         = {Tech. Rep.},
  year         = {2025},
  month        = may,
  url          = {https://www.verizon.com/business/resources/reports/dbir/},
  note         = {Accessed: 2025-11-09}
}

@misc{unb2018Datasets,
	author = {{Canadian Institute for Cybersecurity}},
	title = {{I}{D}{S} 2018 | {D}atasets | {R}esearch | {C}anadian {I}nstitute for {C}ybersecurity | {U}{N}{B} --- unb.ca},
	howpublished = {\url{https://www.unb.ca/cic/datasets/ids-2018.html}},
	year = {},
	note = {[Accessed 09-11-2025]},
}

@misc{opendataRealisticCyber,
	author = {{Amazon Web Services}},
	title = {{A} {R}ealistic {C}yber {D}efense {D}ataset ({C}{S}{E}-{C}{I}{C}-{I}{D}{S}2018) - {R}egistry of {O}pen {D}ata on {A}{W}{S} --- registry.opendata.aws},
	howpublished = {\url{https://registry.opendata.aws/cse-cic-ids2018/}},
	year = {},
	note = {[Accessed 09-11-2025]},
}

@misc{unswUNSWNB15Dataset,
	author = {{UNSW Canberra Cyber}},
	title = {{T}he {U}{N}{S}{W}-{N}{B}15 {D}ataset | {U}{N}{S}{W} {R}esearch --- research.unsw.edu.au},
	howpublished = {\url{https://research.unsw.edu.au/projects/unsw-nb15-dataset}},
	year = {},
	note = {[Accessed 09-11-2025]},
}

@misc{unbApplicationsResearch,
	author = {{Canadian Institute for Cybersecurity}},
	title = {{A}pplications | {R}esearch | {C}anadian {I}nstitute for {C}ybersecurity | {U}{N}{B} --- unb.ca},
	howpublished = {\url{https://www.unb.ca/cic/research/applications.html}},
	year = {},
	note = {[Accessed 09-11-2025]},
}

@article{leevy2020survey,
  title={A survey and analysis of intrusion detection models based on cse-cic-ids2018 big data},
  author={Leevy, Joffrey L and Khoshgoftaar, Taghi M},
  journal={Journal of Big Data},
  volume={7},
  number={1},
  pages={104},
  year={2020},
  publisher={Springer}
}

@article{songma2023optimizing,
  title={Optimizing intrusion detection systems in three phases on the CSE-CIC-IDS-2018 dataset},
  author={Songma, Surasit and Sathuphan, Theera and Pamutha, Thanakorn},
  journal={Computers},
  volume={12},
  number={12},
  pages={245},
  year={2023},
  publisher={MDPI}
}

@article{gama2014survey,
  title={A survey on concept drift adaptation},
  author={Gama, Jo{\~a}o and {\v{Z}}liobait{\.e}, Indr{\.e} and Bifet, Albert and Pechenizkiy, Mykola and Bouchachia, Abdelhamid},
  journal={ACM computing surveys (CSUR)},
  volume={46},
  number={4},
  pages={1--37},
  year={2014},
  publisher={ACM New York, NY, USA}
}

\end{document}